\newif\ifCHANGES
\let\ifCHANGES\iftrue

\documentclass[3p]{elsarticle}

\usepackage[utf8]{inputenc}

\usepackage{etoolbox}
\usepackage{xparse}

\usepackage{microtype}

\usepackage{amsmath}
\usepackage{amssymb}
\usepackage{mathtools}
\usepackage{bm}
\usepackage{upgreek}
\usepackage{physics}
\usepackage{siunitx}

\usepackage{caption}
\usepackage{csquotes}
\usepackage[inline]{enumitem}
\usepackage{graphicx}
\usepackage{xcolor}

\usepackage{booktabs}

\usepackage{todonotes}

\usepackage{acronym}
\usepackage{hyperref}

\biboptions{sort&compress}

\definecolor{edits}{RGB}{220,0,0}
\definecolor{strike}{RGB}{150,50,50}
\ifCHANGES
    \usepackage[normalem]{ulem}
    
    \NewDocumentCommand\STRIKE{+m}{{\color{strike}\sout{#1}}}
\else
    
    \NewDocumentCommand\STRIKE{+m}{}
\fi

\NewDocumentCommand\eg{}{e.\,g.}
\NewDocumentCommand\ie{}{i.\,e.}
\NewDocumentCommand\cf{}{cf.}

\NewDocumentCommand\eqperiod{}{\,\text{.}}
\NewDocumentCommand\eqcomma{}{\,\text{,}}

\NewDocumentCommand\T{}{\mathsf{T}}
\RenewDocumentCommand\vec{m}{\bm{#1}}
\NewDocumentCommand\vunit{m}{\ensuremath{\vu*{e}_{#1}}}
\NewDocumentCommand\mat{m}{\bm{#1}}

\DeclarePairedDelimiterX\set[1]{\{}{\}}
    {#1}

\NewDocumentCommand\Ws{}{\ensuremath{\mathcal{W}_\mathrm{s}}}
\NewDocumentCommand\Wu{}{\ensuremath{\mathcal{W}_\mathrm{u}}}

\NewDocumentCommand\xDS{}{\ensuremath{x^\mathrm{DS}}}
\NewDocumentCommand\xNHIM{}{\ensuremath{x^\ddagger}}

\NewDocumentCommand\Nreact{}{\ensuremath{N}}
\NewDocumentCommand\kAvrg{}{\ensuremath{\bar{k}}}
\NewDocumentCommand\kEnse{}{\ensuremath{k_\mathrm{e}}}
\NewDocumentCommand\kEnseAvrg{}{\ensuremath{\bar{k}_\mathrm{e}}}
\NewDocumentCommand\kMani{}{\ensuremath{k_\mathrm{m}}}
\NewDocumentCommand\kManiAvrg{}{\ensuremath{\bar{k}_\mathrm{m}}}
\NewDocumentCommand\kFloq{}{\ensuremath{\bar{k}_\mathrm{F}}}

\NewDocumentCommand\bS{}{\mathrm{S}}
\NewDocumentCommand\bB{}{\mathrm{B}}
\NewDocumentCommand\bE{}{\mathrm{P}}
\NewDocumentCommand\bM{}{\mathrm{M}}

\NewDocumentCommand\muB{}{\ensuremath{\mu}}
\NewDocumentCommand\muM{}{\ensuremath{\tilde{\mu}}}
\NewDocumentCommand\mass{m}{\ensuremath{M_{#1}}}

\NewDocumentCommand\pos{m}{\ensuremath{\vec{R}_{#1}}}
\NewDocumentCommand\vr{}{\ensuremath{\vec{r}}}
\NewDocumentCommand\vdist{m}{\ensuremath{\vr_{#1}}}
\NewDocumentCommand\dist{m}{\ensuremath{r_{#1}}}
\NewDocumentCommand\Veff{}{\ensuremath{V_\mathrm{eff}}}

\NewDocumentCommand\aDSdot{}{\ensuremath{\dot{\alpha}^\mathrm{DS}}}
\NewDocumentCommand\aDS{}{\ensuremath{\alpha^\mathrm{DS}}}
\NewDocumentCommand\gammaE{}{\ensuremath{\vec{\gamma}^\mathrm{ens}}}
\NewDocumentCommand\gammaS{}{\ensuremath{\vec{\gamma}^\mathrm{s}}}
\NewDocumentCommand\gammaTS{}{\ensuremath{\vec{\gamma}^\ddagger}}
\NewDocumentCommand\gammaU{}{\ensuremath{\vec{\gamma}^\mathrm{u}}}
\NewDocumentCommand\kDS{}{\ensuremath{k^\mathrm{DS}}}
\NewDocumentCommand\kEns{}{\ensuremath{k^\mathrm{ens}}}
\NewDocumentCommand\pxS{}{\ensuremath{p_x^\mathrm{s}}}
\NewDocumentCommand\pxTS{}{\ensuremath{p_x^\ddagger}}
\NewDocumentCommand\pxU{}{\ensuremath{p_x^\mathrm{u}}}
\NewDocumentCommand\pyTS{}{\ensuremath{\vec{p}_y^\ddagger}}
\NewDocumentCommand\xS{}{\ensuremath{x^\mathrm{s}}}
\NewDocumentCommand\xTS{}{\ensuremath{x^\ddagger}}
\NewDocumentCommand\xU{}{\ensuremath{x^\mathrm{u}}}
\NewDocumentCommand\yTS{}{\ensuremath{\vec{y}^\ddagger}}

\DeclareSIUnit\MJupiter{\mathit{M}_J}

\journal{Communications in Nonlinear Science and Numerical Simulation}

\begin{document}

\title{On the stability of satellites at unstable libration points
     of sun--planet--moon systems}

\author[1]{Johannes Reiff}
\author[1]{Jonas Zatsch}
\author[1]{J\"org Main}
\address[1]{
    Institut f\"ur Theoretische Physik I,
    Universit\"at Stuttgart,
    70550 Stuttgart, Germany
}

\author[2,3]{Rigoberto Hernandez\texorpdfstring{\corref{cor}}{}}
\ead{r.hernandez@jhu.edu}
\cortext[cor]{Corresponding author}
\address[2]{
    Department of Chemistry,
    Johns Hopkins University,
    Baltimore, Maryland 21218, United States
}
\address[3]{
    Departments of Chemical \& Biomolecular Engineering,
    and Materials Science and Engineering,
    Johns Hopkins University,
    Baltimore, Maryland 21218, United States
}

\date{\today}

\begin{abstract}
    The five libration points of a sun--planet system
    are stable or unstable fixed positions
    at which satellites or asteroids can remain fixed
    relative to the two orbiting bodies.
    A moon orbiting around the planet
    causes a time-dependent perturbation on the system.
    Here, we address the sense in which invariant structure remains.
    We employ a transition state theory developed previously
    for periodically driven systems with a rank-1 saddle
    in the context of chemical reactions.
    We find that a satellite can be parked
    on a so-called time-periodic transition state
    trajectory---which is an orbit restricted to the vicinity
    of the libration point L2
    for infinitely long time---and
    investigate the stability properties of that orbit.
\end{abstract}


\maketitle


\acrodef{BCM}{binary contraction method}
\acrodef{BCR4BP}{bicircular restricted four-body problem}
\acrodef{CR3BP}{circular restricted three-body problem}
\acrodef{DoF}{degree of freedom}
\acrodefplural{DoF}{degrees of freedom}
\acrodef{DS}{dividing surface}
\acrodef{LMA}{local manifold analysis}
\acrodef{NHIM}{normally hyperbolic invariant manifold}
\acrodef{PES}{potential energy surface}
\acrodef{PSOS}{Poincaré surface of section}
\acrodef{TS}{transition state}
\acrodef{TST}{transition state theory}


\section{Introduction}
\label{sec:intro}

\begin{figure}
    \centering
    \includegraphics[width=\linewidth]{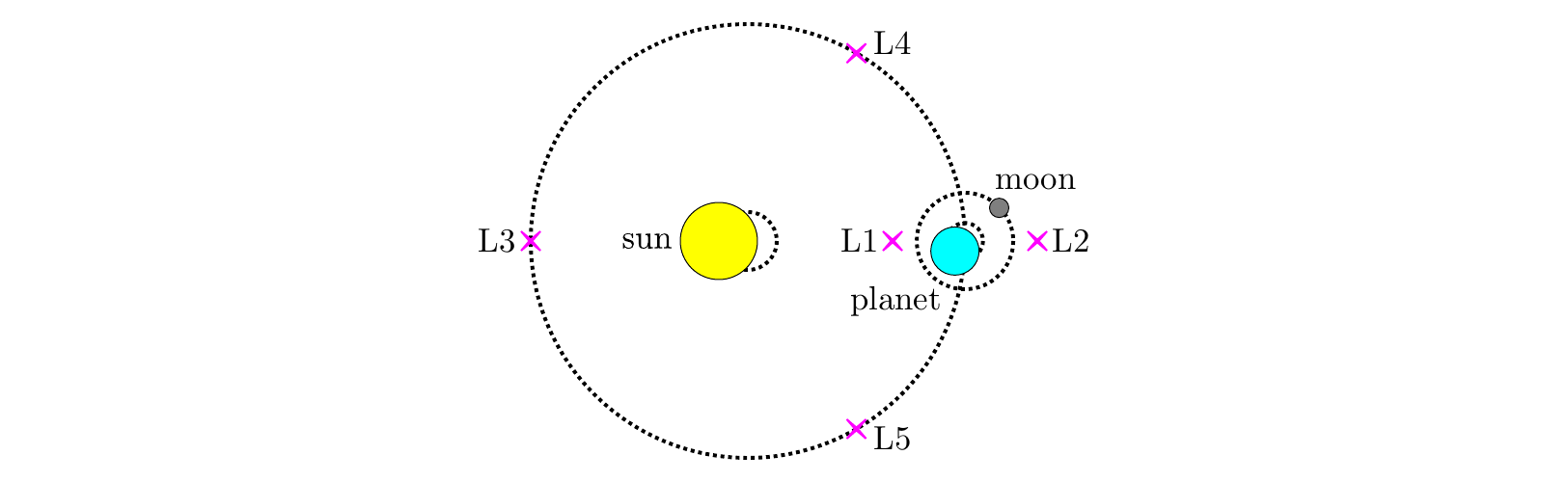}
    \caption{%
        Schematic of a generic solar system in the \acs{BCR4BP}.
        A planet and its moon are on circular orbits about their mutual barycenter.
        This planet--moon barycenter and the sun both
        move around the total center of mass.
        Other bodies in this system%
        ---\eg, satellites or asteroids---%
        are assumed to have negligible mass.
        All orbits are coplanar.
        Without the moon, the system is known to give rise to
        the five libration points, marked L1 to L5, associated with the
        stability of additional objects such as a satellite or asteroid.
        The unstable point L2 is the focus of this work.
        In the special case of our Solar System, these bodies
        refer to the Sun, the Earth and the Moon.
        Here, we consider a broader range of cases
        and list the corresponding parameters
        in Table~\ref{tab:solar_system/params}.}
    \label{fig:solar_system_schematic}
\end{figure}

It is well known that a balance of forces between bodies in space can
lead to stable or unstable fixed points in which a small body, such
as a craft or satellite, experiences no net forces in a particular
moving frame.
The geostationary points arising from the cancellation of
the gravitational force of the Earth and a satellite's centrifugal force
is a nice example.
At this fixed point, the satellite
remains forever fixed above a particular point on the Earth's surface
as both rotate in tandem.
Five fixed points are known to exist in the two-body Sun--Earth system
and are called the libration or Lagrange points
L1 to L5~\cite{Steg1966, Murray2000},
as schematically illustrated in Fig.~\ref{fig:solar_system_schematic}.
The three collinear libration points L1 to L3 are unstable because
small deviations from the exact position increase with time
leading to the decay of the satellite away from them.
However, the other two triangular libration points L4 and L5 can be stable
because of the effects of the Coriolis force,
if the ratio of the masses of the sun and planet is sufficiently
large~\cite{Wintner1941, Steg1966}.

The dynamics of planetary systems becomes much more complicated
as soon as a third primary such as a moon is considered~%
\cite{Cronin1964, Simo1995, Koon2001, Guo2019, Negri2020a}.
In this case,
the stability and position of the libration points in the rotating frame
are influenced by the Moon's gravitational force.
That is, the Moon's rotation around the Earth causes
a time-dependent periodic driving of the satellite
and it is a non-trivial question whether or not the satellite
can still stay forever in the local vicinity of a libration point.
In this paper, we address this question with respect to the
libration point L2.
We characterize the dynamics and stability of a satellite near
this point in a generic sun--planet system
while explicitly considering the effects from the moon's rotation.
The dynamics near the unstable libration points
is determined by a rank-1 saddle of the potential.
To investigate the dynamics close to the saddle we resort to
the use of \ac{TST}~%
\cite{eyring35, wigner37, pitzer, pech81, truh96, peters14a, wiggins16}.
In chemical reactions without driving,
the precise determination of a \ac{DS} separating reactants and products
is key to determining rate constants.
TST rates are given by the particle flux through that surface~\cite{
    truh79, truh85, hynes85b, berne88, nitzan88, rmp90, truhlar91,
    truh2000, KomatsuzakiBerry01a, pollak05a, hern10a, Henkelman2016}.
It has been applied in various fields of physics and chemistry,
including, \eg, atomic physics~\cite{Jaffe00},
solid state physics~\cite{Jacucci1984},
cluster formation~\cite{KomatsuzakiBerry99a, KomatsuzakiBerry02},
diffusion dynamics~\cite{toller, voter02b}, and
cosmology~\cite{Oliveira02}, to name a few.
A version of \ac{TST} has also been implemented to describe the
escape of satellites from planetary neighborhoods in
celestial mechanics~\cite{Jaffe02, Waalkens2005b}.
Here, we use recent advances in \ac{TST} for
driven chemical reactions~\cite{hern19a, hern19e, hern20d, hern20m}
to generalize such a treatment
to include the effects of a moon on the escape of a satellite from L2.

The paper is organized as follows:
In Sec.~\ref{sec:solar_system}, we introduce the formulation of
the \ac{BCR4BP}~\cite{Cronin1964, Simo1995, Koon2001, Guo2019}
used throughout this paper.
In addition to specifying the system, we also
derive the corresponding equations of motion
and give a short review of the libration points.
In Sec.~\ref{sec:tst}, we present a brief introduction to \ac{TST} and
the numerical methods used for the analysis of the phase space structure,
the determination of the periodic libration point orbit,
and the stability analysis of that orbit.
Finally, we show in Sec.~\ref{sec:results} that
the fixed point associated with the L2 evolves into
a periodic libration point orbit
when the moon's effects on the system are explicitly included.
Our central finding is that the decay rates for satellites
on this orbit can be obtained using a transition state theory
framework, and in so doing, revealing the geometric
structure of its dynamics.
Using the calculated decay rates for satellites on this orbit,
the \ac{TST} approach allows for the determination of
the effort needed to actively stabilize satellites near L2
as a time-dependent function of the moon's position.


\section{The model solar system}
\label{sec:solar_system}

We focus on planetary systems with three primary bodies
to illustrate the use of \ac{TST} in celestial mechanics.
Unfortunately, there does not exist a general closed-form solution for such
three-body problems~\cite{Barrow1997a} that would provide
a rigorous benchmark.
We therefore base our calculations on the well-known \ac{BCR4BP}
which models three primaries on fixed circular orbits
in the same two-dimensional plane~%
\cite{Cronin1964, Simo1995, Koon2001, Guo2019}.
The two lighter primaries---referred to as planet and moon---%
orbit their mutual barycenter
in an inner nested two-body problem.
In turn, this planet--moon barycenter and the heaviest primary---the sun---%
orbit the total system's center of mass
as shown in Fig.~\ref{fig:solar_system_schematic}
Although this center of mass is
itself a barycenter for the whole system, we will avoid
referring it as such through the text to avoid confusion.
In the outer nested two-body problem,
the planet and moon are treated as a single body located at their barycenter.
The dynamics of the probe body (for example an asteroid or satellite)
is governed by the primaries' gravitational forces.
Its mass is assumed to be so small that its gravitational force on other bodies
can be neglected.


\subsection{Libration points in the sun--planet system}
\label{sec:solar_system/libration_points}

A system consisting of two bodies interacting via gravitational forces%
---in our case a sun and a planet---%
features five points where the gravitational forces and the centrifugal force
on a probe mass nullify each other.
These equilibrium points are called
the libration or Lagrange points L1 to L5.
Figure~\ref{fig:solar_system_schematic} schematically shows
the position of L1 to L5 relative to the celestial bodies.
The libration points rotate with the same frequency as the primaries,
and therefore the probe mass' position relative to the bodies is constant.
This makes them very interesting for space exploration and research~%
\cite{Fleck1997a, Armano2016a, Kraft2017a}.
Adding a lighter third primary---\ie, a moon orbiting the planet---%
causes a time-dependent perturbation of the libration points.
See Sec.~\ref{sec:results/dynamics} for a more detailed discussion.

The libration points differ by their stability properties.
The potentials near L1 to L3 are rank-1 saddles
with one unstable \ac{DoF} in the radial direction
and one stable \acp{DoF} perpendicular to the radial direction.
Trajectories near L1 to L3 are therefore \emph{unstable}.
The libration points L4 and L5 show as rank-2 saddles
with two unstable directions.
Contrary to first intuition, however,
trajectories near L4 and L5 are \emph{Lyapunov stable}
for sufficiently large sun--planet mass ratios~\cite{Wintner1941, Steg1966}
because of the contributions from the velocity-dependent Coriolis force~%
\cite{Wintner1941, Bhatnagar1978a}.
However, trajectories near the collinear libration points L1 to L3
are not stabilized in similar fashion
despite the effects from the Coriolis force.

In this paper we will focus on the stability of satellites near L2
and the influence of a moon thereupon.
In our Solar System, L2 is especially interesting for astrophysics
since it is located far enough from Sun, Earth, and Moon to minimize noise
but still close enough to Earth for communication.
It has been the target of multiple past and present space missions, \eg,
the \emph{Wilkinson Microwave Anisotropy Probe}~\cite{Bennett2003},
the \emph{James Webb Space Telescope}~\cite{Harwit2004},
the \emph{Herschel} mission~\cite{Gardner2006},
and the \emph{Planck} spacecraft~\cite{PlanckCollaboration2014}.
In addition, it features a rank-1 saddle which can be readily described
using the framework of \ac{TST} as summarized in Sec.~\ref{sec:tst}.


\subsection{Parameters and model variants}
\label{sec:solar_system/params}

\begin{table}
    \centering
    \caption{%
        Summary of the configuration parameters and derived values
        used in the strong-driving and Solar System models considered here.
        We use dimensionless units derived from a synodic coordinate system
        with respect to the sun and the planet--moon barycenter.
        The sun--barycenter distance, sun--barycenter angular frequency,
        and total mass of all primaries
        are used as units for length, frequency, and mass, respectively.
        The satellite is assumed to be an infinitesimal mass point.
        Its mass gets canceled from the equations of motion,
        and there is no need to specify it.
        See Sec.~\ref{sec:solar_system/eom} for further details.}
   \begin{tabular}{l l S[table-format=1.2] S[table-format=2.3e-1]}
        \toprule
        description & symbol & {strong-driving model} & {Solar System model} \\
        \midrule
        planet--moon distance    & $a$                              & 0.1  &  2.570e-3 \\
        primary mass parameter   & $\muB$                           & 0.1  &  3.040e-6 \\
        secondary mass parameter & $\muM$                           & 0.1  &  1.215e-2 \\
        \midrule
        sun mass                 & $\mass{\bS} = 1 - \muB$          & 0.9  &  0.999997 \\
        barycenter mass          & $\mass{\bB} = \muB$              & 0.1  &  3.040e-6 \\
        planet mass              & $\mass{\bE} = \muB (1 - \muM)$   & 0.09 &  3.003e-6 \\
        moon mass                & $\mass{\bM} = \muB \muM$         & 0.01 &  3.695e-8 \\
        synodic moon frequency   & $\omega = \sqrt{\muB / a^3} - 1$ & 9    & 12.387    \\
        \bottomrule
    \end{tabular}
    \label{tab:solar_system/params}
\end{table}

We use dimensionless units derived from a synodic reference frame
with respect to the sun and the planet--moon barycenter.
The sun--barycenter distance, the sun--barycenter angular frequency,
and the total mass of all primaries
act as units for length, frequency (inverse time),
and mass, respectively.
The gravitational constant follows as $G = 1$.
In this coordinate system, the total center of mass is at the origin
with the sun and the barycenter located at
$(-\muB, 0)^\T$ and $(1 - \muB, 0)^\T$, respectively.
The primary mass parameter \muB\ is defined as
the barycenter mass (\ie, sum of planet and moon mass)
in units of the system's total mass.
The planet and moon orbit with distance $a$ and angular frequency $\omega$
around the barycenter.
The secondary mass parameter \muM\ is defined analogously
as the moon mass divided by the barycenter mass,
\cf\ Table~\ref{tab:solar_system/params}.

The \ac{BCR4BP} is only an approximation to the real dynamics.
We can recover strict Newtonian motion of the primaries
by removing the moon ($\muM = 0$).
This limiting case is equivalent to the \ac{CR3BP}
and will be referred to as the time-invariant or \emph{static} system
because the resulting potential is time-independent in the synodic frame.
In contrast, models with $\muM \ne 0$ will be called \emph{driven}
as the moon perturbs the potential even in the synodic frame.

In this work, we consider two parameterizations of the \ac{BCR4BP}
listed in Table~\ref{tab:solar_system/params}:
\begin{enumerate}[label=(\roman*)]
\item
    In the \emph{strong-driving model},
    the mass parameters are taken as $\muB = \muM = \num{0.1}$
    and the planet--moon distance as $a = \num{0.1}$.
    The chosen values lead to a stronger perturbation of L2
    and allow for easier visualization.
    This parameter set is characteristic of
    the more extreme mass ratios that have been observed
    in extrasolar systems.
    An example is the brown dwarf 2M1207A
    and its companion planet 2M1207b~\cite{Ricci2017}
    located in the constellation Centaurus
    with masses around \SI{60}{\MJupiter} and \SI{5}{\MJupiter}
    (\si{\MJupiter} being the Jovian mass), respectively.
    Therein the mass parameter is $\muB \approx \num{0.077}$,
    and is comparable to the value in our parameter set.
    The corresponding effective potential
    is shown in Fig.~\ref{fig:energy_surface}.
\item
    In our \emph{Solar System}, the masses of the Sun,
    the Earth, the Moon, and the relative distances between them
    are well known.
    It provides a comparison to the strong-driving model, and
    demonstrates the applicability of our methods.
\end{enumerate}


\subsection{Potential and equations of motion}
\label{sec:solar_system/eom}

\begin{figure}
    \centering
    \includegraphics[width=\linewidth]{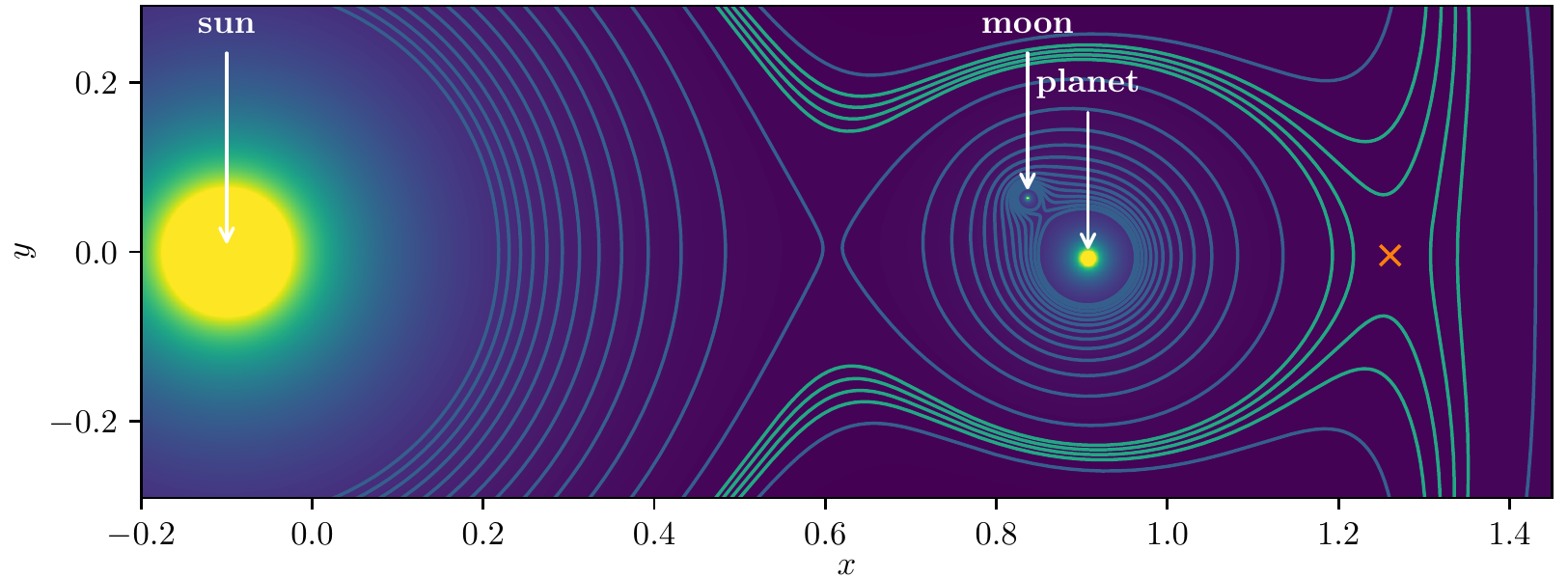}
    \caption{%
        Effective potential $\Veff(x, y, t)$
        according to Eq.~\eqref{eq:solar_system/effective_potential}
        at $t = 3 T / 8$ for the strong-driving model.
        The color map is capped at $\Veff = \num{-12}$
        since the potential is divergent at the positions
        of the system's sun, planet, and moon (filled yellow circles).
        A cross indicates the dynamical position of
        the saddle point near L2.
        We caution the reader that this point does not
        necessarily correspond to an equilibrium position
        due to the potential's time dependence.
        More equipotential lines are shown near L2
        to elucidate the saddle structure.}
    \label{fig:energy_surface}
\end{figure}

The bodies have normalized masses
\begin{equation}
    \mass{\bS} = 1 - \mu
    \qc
    \mass{\bB} = \mu
    \qc
    \mass{\bE} = \mu (1 - \tilde{\mu})
    \qc\qand*
    \mass{\bM} = \mu \tilde{\mu}
\end{equation}
where
\begin{equation}
    0 \le \mu, \tilde{\mu} \le 1
    \qand
    \mass{\bS} + \mass{\bB} = \mass{\bS} + \mass{\bE} + \mass{\bM} = 1
    \eqperiod
\end{equation}
They are located at positions
\begin{equation}
    \pos{\bS} = -\mu \vunit{x}
    \qc
    \pos{\bB} = (1 - \mu) \vunit{x}
    \qc
    \pos{\bE} = \pos{\bB} - a \tilde{\mu} \vunit{\theta}
    \qc\qand*
    \pos{\bM} = \pos{\bB} + a (1 - \tilde{\mu}) \vunit{\theta}
\end{equation}
where
\begin{equation}
    \vunit{\theta} = \mqty(\cos\theta \\ \sin\theta)
    \qq{with}
    \theta = \omega t
    \qand
    \omega = \sqrt{\mu / a^3} - 1
\end{equation}
in the synodic coordinate system.
Using the position of the satellite \vr\ relative to the primaries
\begin{equation}
    \vdist{j} = \vr - \pos{j}
    \qfor j \in \set{\bS, \bE, \bM}
\end{equation}
the effective potential can be written as
\begin{equation}
    \label{eq:solar_system/effective_potential}
    -V_\mathrm{eff}
    = \Omega
    = \frac{r^2}{2}
        + \frac{1 - \mu}{\dist{\bS}}
        + \frac{\mu (1 - \tilde{\mu})}{\dist{\bE}}
        + \frac{\mu \tilde{\mu}}{\dist{\bM}}
    \eqperiod
\end{equation}
This potential for the parameters of the strong-driving model
is shown in Fig.~\ref{fig:energy_surface}.

The equations of motion for $\vr = (x, y)^\T$
can be derived from the effective potential as
\begin{equation}
    \label{eq:solar_system/eom}
    \dot{p}_x = \ddot{x} - 2 \dot{y} = \pdv{\Omega}{x}
    \qand
    \dot{p}_y = \ddot{y} + 2 \dot{x} = \pdv{\Omega}{y}
\end{equation}
where
\begin{equation}
    \pdv{\Omega}{j}
        = j - \sum_{k \in \set{\bS, \bE, \bM}}
            \frac{\mass{k}}{\dist{k}^3} \vdist{k} \vdot \vunit{j}
    \qfor j \in \set{x, y}
    \eqperiod
\end{equation}
The terms $-2 \dot{y}$ and $+2 \dot{x}$ in Eqs.~\eqref{eq:solar_system/eom}
account for the Coriolis force.


\section{Transition state theory}
\label{sec:tst}

The sun--planet--moon systems---described by
the effective potential~\eqref{eq:solar_system/effective_potential}---%
features a rank-1 saddle point at L2,
as shown in Fig.~\ref{fig:energy_surface},
which can act as bottleneck of the dynamics.
This makes it a prime candidate for the application
of \ac{TST} models.
(See, e.g., Refs.~\cite{Jaffe02, Astakhov2003, Waalkens2005b}.)
In typical scenarios for a chemical reaction,
a rank-1 saddle point separates reactants from products, and
can be used to characterize the flux and associated reaction rate.
In the planetary system, reactants roughly correspond to satellites or asteroids
inside the planet's Hill sphere~\cite{Murray2000, Astakhov2003},
and the reaction corresponds to the escape from this sphere,
or vice versa.
The unstable states between reactants and products form the \ac{TS}.
Thus geometrically, the \ac{TS} is a boundary
(or \ac{DS}) between the reactant and product regions
which in the current context corresponds to the
\ac{DS} between incoming or outgoing directions
of the satellite.
In planetary systems, this includes all libration point orbits.
The reaction coordinate~$x$ (in the rotating frame)
describes the saddle's unstable direction
and indicates the progress of the reaction, see Fig.~\ref{fig:tst_basics}(a).
The stable directions are referred to as orthogonal modes~$\vec{y}$.


\subsection{Geometry of the transition state}
\label{sec:tst/geometry}

\begin{figure}
    \centering
    \includegraphics[width=\linewidth]{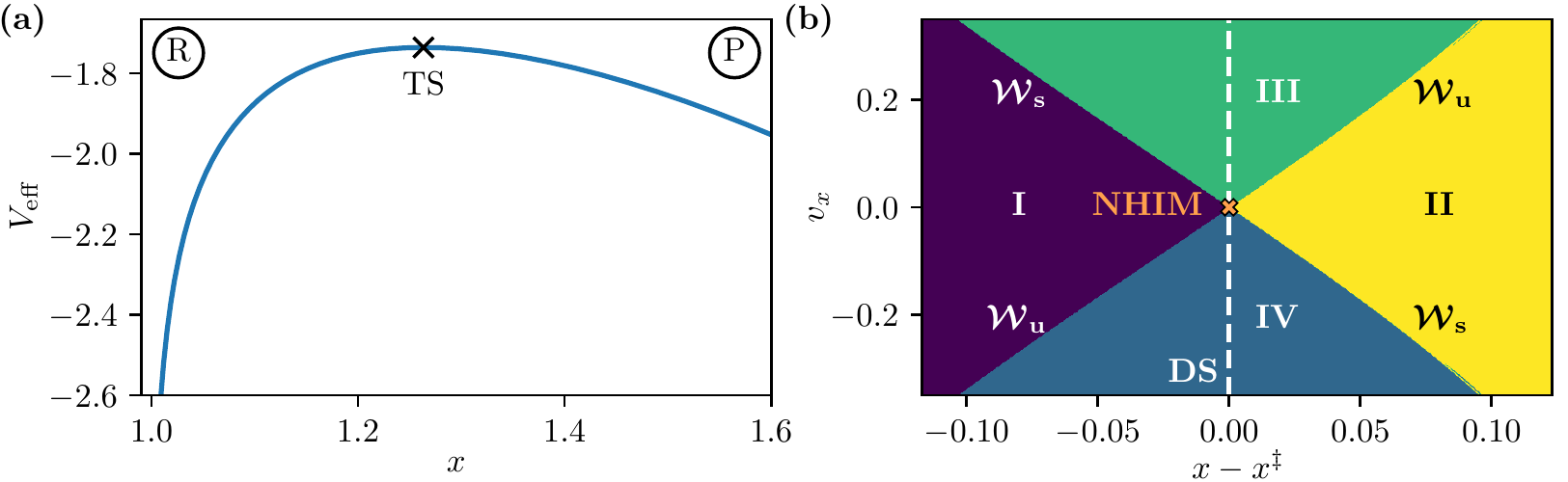}
    \caption{%
        Position space and phase space at $t_0 = 0$ near L2 of
        the strong-driving model defined in Table~\ref{tab:solar_system/params}.
        (a)~A one-dimensional cut of
        the effective potential $\Veff(x, y = 0)$ along $x$.
        Reactants (R) and Products (P) are separated by a saddle in \Veff.
        (b)~The corresponding cut of
        the structures in phase space $(x, y = 0, v_x, v_y = 0)$
        near the \acs{NHIM} at $\xNHIM(y = 0, v_y = 0) \approx \num{1.26030}$.
        The \acs{NHIM}'s stable \Ws\ and unstable \Wu\ manifolds
        divide the phase space into
        four nonreactive (I and II) and reactive (III and IV) regions.
        The \acs{NHIM}
        acts as an anchor for the \acs{DS}.}
    \label{fig:tst_basics}
\end{figure}

While it is straightforward to define a \ac{DS} for a static 1-\ac{DoF} system,
differentiating between reactants and products
in higher-dimensional or driven systems
requires a more sophisticated mathematical framework
though the analogy to the sun--planet--moon system remains.

We start by looking at the phase space near the saddle point.
Depending on where they come from when propagating backward in time
and where they go to when propagating forward in time,
initial conditions of trajectories can be classified as either
\begin{enumerate*}[label=(\Roman*)]
    \item nonreactive reactants,
    \item nonreactive products,
    \item reactive reactants, or
    \item reactive products.
\end{enumerate*}
In the context of libration points,
the terms reactive and nonreactive can be reframed as
\emph{transit} and \emph{nontransit} orbits,
respectively~\cite{Koon2000a, Koon2001}.
The four regions are separated by critical trajectories
forming the stable and unstable manifolds,
as shown in Fig.~\ref{fig:tst_basics}(b).
Trajectories on the stable manifold \Ws\
are bound to the saddle's vicinity for $t \to \infty$
while trajectories on the unstable manifold \Wu\ are bound for $t \to -\infty$.
They play an important role in the determination of
efficient transfer orbits for spacecrafts~%
\cite{Koon2000a, Koon2001, Howell2001a, Gomez2005a, Guo2019}.
These manifolds are associated with
the \ac{NHIM}~%
\cite{Lichtenberg82, hern93b, hern93c, Simo1995, Ott2002a, wiggins2013normally},
where trajectories are bound both forward and backward in time.
The \ac{NHIM} is located at
the intersection of the stable and unstable manifolds' closures.
It is this object that trajectories on \Ws\ and \Wu\ converge towards
for $t \to +\infty$ and $-\infty$, respectively.

The \ac{NHIM} plays an important role in determining
whether a given state is considered a reactant or product---%
viz.\ pre-- or post-- escape in the present
celestial models~\cite{Jaffe02, Astakhov2003, Waalkens2005b}.
As shown in previous work~\cite{wiggins01, Uzer02, hern17h, hern18c, hern19a},
a (locally) recrossing-free codimension-$1$ \ac{DS} can often be constructed
by attaching it to the codimension-$2$ \ac{NHIM}.
In the simplest case, this is done by extending the \ac{NHIM} in $v_x$ direction,
as shown in Fig.~\ref{fig:tst_basics}(b).
The \ac{DS} then divides the phase space into a reactant and a product half.
This construction even works when
the saddle is subjected to time-dependent driving.
In this case, the \ac{NHIM} detaches from the position of the saddle point.

The \ac{NHIM} can be obtained using several
perturbative~\cite{KomatsuzakiBerry01a, KomatsuzakiBerry02}
and direct~\cite{hern15e, wiggins16, hern17h} methods,
but we have found that the \ac{BCM} introduced in Ref.~\cite{hern18g}
is effective and efficient for systems such as the one addressed here.
The algorithm in the \ac{BCM} is initialized by defining a quadrangle
with each of its corners lying exclusively
within one of the four regions shown in Fig.~\ref{fig:tst_basics}(b).
In each iterative step, we first determine an edge's midpoint.
Then, the adjacent corner corresponding to the same region as that midpoint
is moved to the midpoint's position.
By repeating this interleaved bisection procedure in turn for all edges,
the quadrangle successively contracts and converges towards the \ac{NHIM}.


\subsection{Instantaneous decay rates}
\label{sec:tst/instant_rate}

The stability of the \ac{TS} near threshold energy can be quantified
via the decay of trajectories near the \ac{NHIM}.
This is equivalent to
the rate at which an ensemble of reactants%
---\ie, asteroids or satellites on one side of the \ac{DS}---moves
through the \ac{DS}.
The decay of this reactant population is exponentially fast
because of the hyperbolic nature of the \ac{NHIM}.
We can therefore define an instantaneous decay rate $k(t)$
via the differential equation~\cite{Lehmann00a, hern19e, hern20m, hern21a}
\begin{equation}
    \label{eq:tst/decay}
    k(t) = -\frac{\dot{\Nreact}(t)}{\Nreact(t)}
    \eqcomma
\end{equation}
where $\Nreact(t)$ is the time-dependent size of the reactant population.
Compared to other measures of instability,
instantaneous decay rates have the advantage that
they can be evaluated for a specific point in time.
Floquet multipliers and Lyapunov exponents, for instance,
only yield average values for a whole period and long-term limits, respectively.

\begin{figure}
    \centering
    \includegraphics[width=0.5\linewidth]{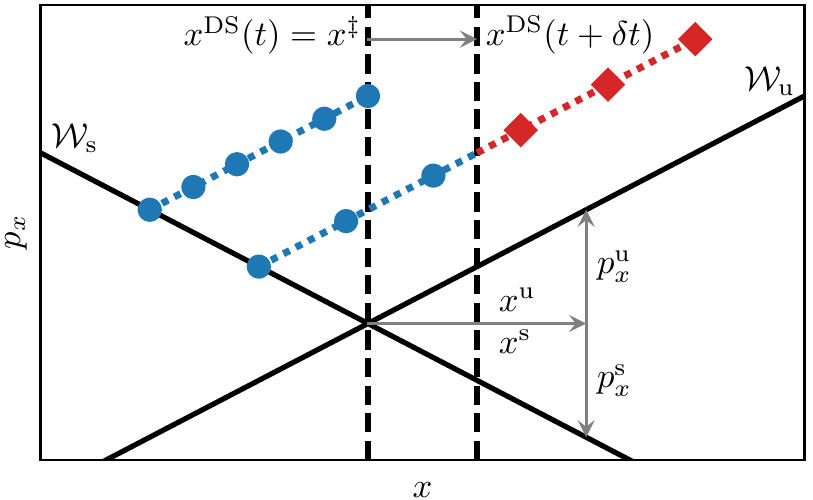}
    \caption{%
        Schematic of instantaneous rate calculations.
        Initially, an equidistant ensemble connecting \Ws\ on the reactant side
        with the \acs{DS} parallel to \Wu\ is generated.
        Upon time propagation, parts of the ensemble react through the \acs{DS}.
        The resulting ensemble is still equidistant,
        parallel to \Wu, and connected to \Ws.}
    \label{fig:ensemble}
\end{figure}

The ensemble method~\cite{hern19e, hern20m, hern21a}
can be used to obtain
the instantaneous decay rate
of a reactant population close to the \ac{NHIM},
which anchors the \ac{TS}.
A homogeneous and linear ensemble of \Nreact\ reactive trajectories
is initialized on the reactant side of the full phase space.
Specifically, they are placed on an $(x, p_x)$-cross sectional surface
at a small distance $\delta x$
from a given position $(\xTS(t), \yTS(t), \pxTS(t), \pyTS(t))^\T$
of an arbitrarily chosen trajectory on the \ac{NHIM}
(see upper bullets in Fig.~\ref{fig:ensemble}).
After propagating this ensemble for a time $\delta t$,
a subdomain will have pierced the \ac{DS} and entered the product side
(diamonds in Fig.~\ref{fig:ensemble})
while the remainder will still be located on the reactant side
(lower bullets in Fig.~\ref{fig:ensemble}).
As the \ac{DS} is non-recrossing,
the resulting decrease of the reactant population
in a close neighborhood of the \ac{NHIM}
(\cf\ Eq.~\eqref{eq:tst/decay})
is associated with a decay rate
\begin{equation}
    \label{eq:tst/ensemble_rate}
    \kEnse(\vec{y}, \vec{p}_y, t)
    = -\dv{t} \ln[\Nreact(\vec{y}, \vec{p}_y, t)]
    \eqperiod
\end{equation}
It is referred to as the \emph{instantaneous ensemble rate} to emphasize that
it is obtained by propagating an ensemble of reactive trajectories
according to the equations of motion.
In doing so, the \ac{DS} is computed
individually for each trajectory of the ensemble and each time step.

The ensemble can be propagated not just for a small time step $\delta t$
but for longer times when computing the time-dependent ensemble rate
according to Eq.~\eqref{eq:tst/ensemble_rate}.
As the reactant population decreases exponentially
when propagating the ensemble,
a new ensemble can be initialized close to
the corresponding point of the respective trajectory on the \ac{NHIM}
after an appropriately chosen propagation time.
Such technical details are discussed in Ref.~\cite{hern19e}.
Although the implementation of the ensemble method is straightforward,
it can be numerically expensive
because the ensemble consists of many trajectories and
the \ac{DS} is obtained individually for each reactive trajectory
using the \ac{BCM}~\cite{hern18g}.

The accuracy of the ensemble method decay rates decreases with time
because of the reduction of the density of reactants.
However, the inaccuracy can be reduced at the cost
of larger computational expense by starting bigger ensembles
or starting the ensembles staggered.
Alternatively,
we can avoid most of the expensive particle propagation of the ensemble method
by leveraging the geometry of the stable and unstable manifolds
to effectively describe
the linearized dynamics near the \ac{NHIM} (see \ref{sec:jacobian}).
The resulting \ac{LMA} method~\cite{hern19e}
can thus be seen as an extension to the ensemble method
with the difference that the time when trajectories have reached the \ac{DS}
can now be determined analytically through the linearization
thereby avoiding a costly numerical integration.
As a result,
the computational effort required to calculate instantaneous decay rates
is significantly reduced
while numerical precision is simultaneously enhanced.
For reference, calculating $k$ for a single point on the \ac{NHIM}
(optimized for maximal precision, including two \ac{BCM} calls)
took about \SI{15}{\s} on an Intel i5-3470 CPU at \SI{3.2}{\GHz}
for the system considered here.
Lower-precision results can be obtained significantly faster.
Additionally,
these calculations can be trivially parallelized for entire trajectories.
Compared to previous publications, the \ac{LMA} method was extended to
account for velocity-dependent forces (\eg\ the Coriolis force)
that arise from the rotating bodies in the sun--planet--moon system.
The details can be found in \ref{sec:lma}.


\subsection{Floquet rate method}
\label{sec:tst/floquet}

The \ac{LMA} is significantly cheaper to evaluate
than a full ensemble propagation.
The numerical effort required, however, can still render it impractical
when analyzing a large number of trajectories.
When only time-averaged rate constants of periodic orbits are sought-after,
a more tractable approach is provided by the Floquet rate method
introduced in Ref.~\cite{hern14f}.
It determines the rate constant
\begin{equation}
    \kFloq = \mu_\mathrm{u} - \mu_\mathrm{s}
\end{equation}
for an orbit on the \ac{NHIM} with period $T$
as the difference between the Floquet exponents
\begin{equation}
    \mu_\mathrm{u, s} = \frac{1}{T} \ln|m_\mathrm{u, s}|
    \eqperiod
\end{equation}
The Floquet multipliers $m_\mathrm{u, s}$ are
the eigenvalues of the monodromy matrix, \ie,
of the fundamental matrix $\mat{\sigma}(t; t_0)$ evaluated at time $t = t_0 + T$.
The fundamental matrix is calculated by solving the differential equation
\begin{equation}
    \dot{\mat{\sigma}}(t; t_0) = \mat{J}(t) \mat{\sigma}(t; t_0)
    \qq{with}
    \mat{\sigma}(t_0; t_0) = \mat{1}_{2 d}
\end{equation}
and where $\mat{J}$ is the Jacobian of the equations of motion
(\cf\ \ref{sec:jacobian}),
$d$ is the number of \acp{DoF},
and $\mat{1}_{2 d}$ is the $(2 d \times 2 d)$-dimensional identity matrix.
Evaluating a full trajectory with this method
takes on the order of \SI{10}{\ms} on an Intel i5-3470 CPU at \SI{3.2}{\GHz}.


\section{Results and discussion}
\label{sec:results}

In this section we mainly focus on the dynamics on the \ac{NHIM}
and the rate of instability that can be derived from it.
Since the \ac{NHIM} is a hyperbolic subspace,
trajectories need to be stabilized to not deviate numerically.
This is achieved by periodically projecting satellites back using the \ac{BCM}
(\cf\ Sec.~\ref{sec:tst/geometry}).
When this is done frequently enough,
errors introduced by this projection are negligible.


\subsection{Dynamics on the NHIM}
\label{sec:results/dynamics}

\begin{figure}
    \centering
    \includegraphics[width=\linewidth]{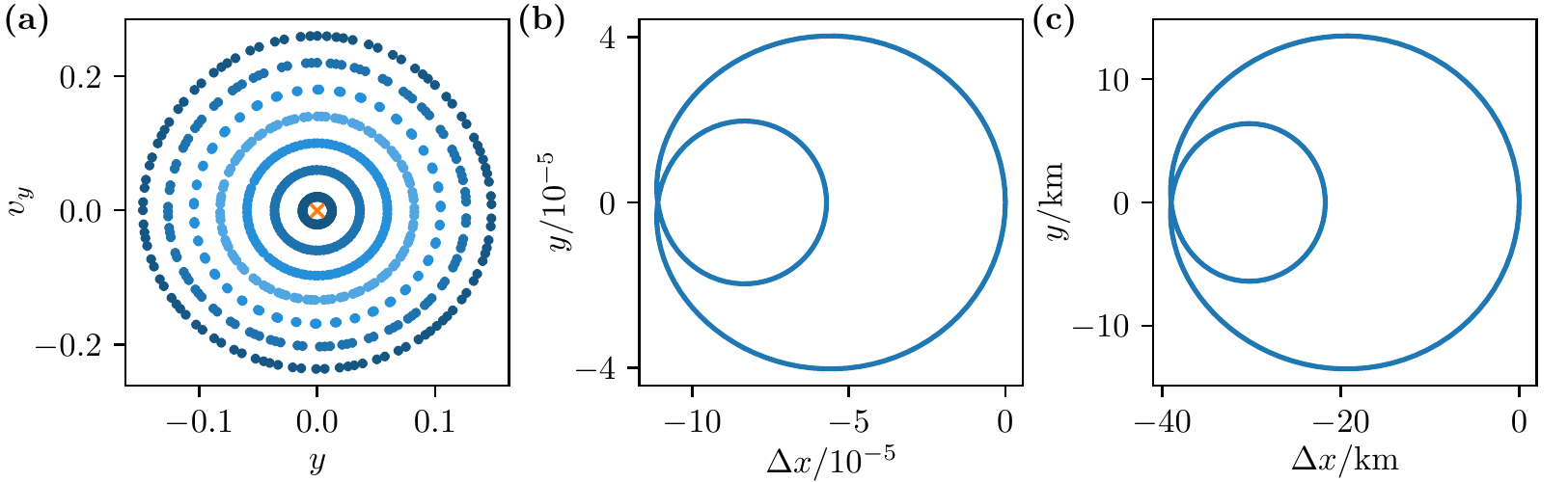}
    \caption{%
        (a)~Stroboscopic \acs{PSOS} of the strong-driving model
        with \num{7} trajectories
        propagated over \num{100} periods on the \acs{NHIM}
        starting at initial time $t_0 = \num{0}$ (blue circles).
        The system exhibits regular dynamics as shown by the concentric tori.
        In its center, a fixed point (orange cross) reveals
        the existence of a periodic orbit with the system's period $T$.
        (b)~The periodic orbit from (a) in position space.
        Positions $\Delta x$ are relative
        to the position $\vec{\gamma}(t_0) \approx
        (\num{1.26047}, \num{0}, \num{0}, \num{-7.6e-4})^\T$.
        (c)~The analogous periodic orbit for the Solar System model.}
    \label{fig:psos}
\end{figure}

The dynamics on the \ac{NHIM} of the periodically driven model system
takes place in an effectively two-dimensional phase space.
An established method for the visualization of such systems
is the stroboscopic map,
a special case of the \ac{PSOS}.
We propagate satellites on the \ac{NHIM}.
Instead of recording the whole trajectory, however,
we only record one point every period $T = 2 \pi / \omega$.
Trajectories with period $T$ therefore manifest as fixed points in the \ac{PSOS}.

Figure~\ref{fig:psos}(a) shows the stroboscopic \ac{PSOS}
of the driven model system's \ac{NHIM} around L2.
One can see concentric toroidal structures, suggesting regular behavior.
Trajectories associated with these tori are mostly quasi-periodic.
The tori surround an elliptic fixed point,
whose associated periodic trajectory is shown in Fig.~\ref{fig:psos}(b).
This orbit has the same period as the moon's synodic orbit.
It can be seen as the generalization of the static L2 point from the \ac{CR3BP}.
Similar observations have been made before,
\eg, in Ref.~\cite{Simo1995} for the triangular libration points.
In the following, we will therefore refer to this trajectory
as the \emph{L2 orbit}.
It has been shown previously~\cite{hern19e, hern20m}
that elliptic fixed points
are typically associated with extrema in the averaged decay rate.
Consequently, there
exists a region around the fixed points with similar decay rates.
Thus to characterize the escape of satellites
in the present time-dependent system,
we must first determine the L2 orbit and then
calculate the decay from it as we do in the following subsection.

The analogous L2 orbit for the Solar System model
is shown in Fig.~\ref{fig:psos}(c).
It exhibits the same structure as the strong-driving model,
but with a much lower orbit diameter of
$\Delta x \approx \num{2.6e-7}$ simulation units.


\subsection{Decay rates}
\label{sec:results/rates}

At this point,
we are able to analyze the stability of satellites close to the L2 in our system.
For this purpose,
we calculate decay rates of satellites following the L2 orbit
with the three methods presented in
Secs.~\ref{sec:tst/instant_rate} and~\ref{sec:tst/floquet}.
The ensemble method has been configured to
propagate \num{32} ensembles of \num{1024} satellites each.
The ensembles are initialized at a maximum distance of
$\delta x = \num{2e-4}$ from the L2 orbit.
We use $\delta x = \num{e-5}$ when determining
the slopes of the stable and unstable manifolds for the \ac{LMA}.

\begin{figure}
    \centering
    \includegraphics[width=\linewidth]{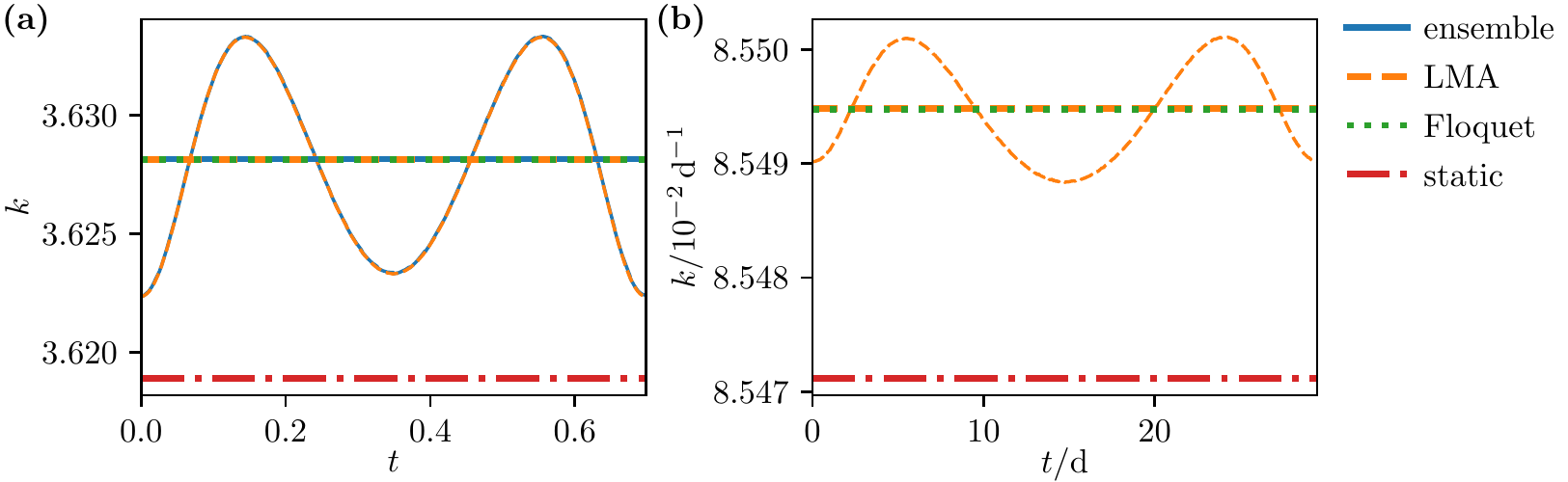}
    \caption{%
        Instantaneous rates $k$ as a function of time $t$
        for the periodic trajectories in Fig.~\ref{fig:psos}.
        Results for the strong-driving model are shown in panel~(a)
        while the Solar System model is represented in panel~(b).
        The instantaneous rate $k(t)$ (thin lines) and
        time average \kAvrg\ (thick lines) are shown for
        the ensemble method \kEnse\ [solid, panel~(a) only] and
        the \acs{LMA} method \kMani\ (dashed).
        They are in agreement with each other, and with the
        corresponding Floquet rate \kFloq\ (thick dotted).
        For comparison, \kFloq\ of the corresponding static systems,
        in which the moon is effectively removed by setting $\muM = 0$,
        is shown with a dash-dotted line.}
    \label{fig:rates}
\end{figure}

The instability of the L2 orbit as determined by the decay rate
is shown in Fig.~\ref{fig:rates}.
Numerical values for the averaged rate constants
are summarized in Table~\ref{tab:results/rates}.
As can be seen, all methods are in excellent agreement.
The relative deviation between mean rates in the driven system is
\begin{equation}
    \abs{\frac{\kAvrg_a - \kAvrg_b}{\kAvrg_b}} < \num{e-5}
    \qq{where}
    \kAvrg_a, \kAvrg_b \in \set*{\kEnseAvrg, \kManiAvrg, \kFloq}
    \eqperiod
\end{equation}
This confirms the accuracy of the results
since all three methods have vastly different numerical implementations.

\begin{table}
    \centering
    \caption{%
        Numerical values for the averaged rate constants
        shown in Fig.~\ref{fig:rates}
        for the corresponding L2 orbits shown in
        Figs.~\ref{fig:psos}(b) and ~\ref{fig:psos}(c).
        The ensemble propagation rate for the Solar System model is not listed
        because it was not calculated so as to avoid the unnecessary
        computational expense.}
    \begin{tabular}{l l S[table-format=1.6] S[table-format=1.6e-1]}
        \toprule
        description & symbol & {strong-driving model} & {Solar System model [\si{\per\day}]} \\
        \midrule
        ensemble propagation       & \kEnseAvrg & 3.628145 &             \\
        local manifold analysis    & \kManiAvrg & 3.628115 & 8.549482e-2 \\
        Floquet stability analysis & \kFloq     & 3.628116 & 8.549479e-2 \\
        static system ($\muM = 0$) & \kFloq     & 3.618910 & 8.547117e-2 \\
        \bottomrule
    \end{tabular}
    \label{tab:results/rates}
\end{table}

We show results for both
the L2 orbit of the driven system (with moon, $\muM > 0$)
and the L2 point of the static system (without moon, $\muM = 0$)
as defined in Sec.~\ref{sec:solar_system}.
In the latter case, we restrict our analysis to the Floquet rate method
as the instantaneous rate in static systems is constant in time
and does not yield additional information.

In the comparison to
the static system
($\muM = 0$) in Fig.~\ref{fig:rates}(a),
we find a \SI[retain-explicit-plus]{+0.3}{\percent} increase
in the average decay rate of the driven system.
That is, the moon generally
destabilizes satellites close to the L2 orbit.
During one period of the orbit,
L2 is most stable around $t \bmod T \in \set{0, T / 2}$,
\ie, when the planet, moon and L2 are collinear.
At these points in time,
either the planet or the moon are at their point of minimal distance from L2,
therefore exerting the strongest forces on satellites near L2.
Stronger forces push the position of L2 further out.
This is consistent with the course of
the L2 orbit shown in Fig.~\ref{fig:psos}(b),
which reaches maxima in $x$ around $t \bmod T \in \set{0, T / 2}$.
The lower decay rates entail that
satellites near but not on the \ac{NHIM} depart slower from L2.
Consequently, fewer course corrections are needed around these points in time.

The limiting case that merges the planet and moon
into a single effective body
can be achieved by setting $\muM = 0$.
This keeps the system's total mass fixed
and only changes the mass distribution between planet and moon.
Alternatively, we can consider the case in which
the sun--planet mass ratio $\mass{\bS} / \mass{\bE}$ is kept fixed
by setting the mass parameter $\muB = \num{0.09} / \num{0.99}$.
The resulting Floquet rate constant now reads $\kFloq \approx \num{3.665}$.
In this case, the addition of a moon leads to
a \SI{-1.0}{\percent} decrease of the rate constant.
Hence, the effect of the moon is similar to
a static system with an increased planet mass.
The resulting stronger forces being exerted on satellites
push the L2 further out.

The calculations of the \ac{LMA} rates and the Floquet rate constants
were also performed for the Solar System model.
The results, shown in Fig.~\ref{fig:rates}(b),
are qualitatively identical to those of the strong-driving model
discussed above.
The decay rate oscillations
and the relative deviation between the static and the driven systems,
however, are quantitatively smaller by an order of magnitude.


\section{Conclusion and outlook}
\label{sec:conclusion}

In this work, we have demonstrated
that the methods from \ac{TST} can be used
to analyze the dynamics of satellites in planetary systems.
Specifically, we analyzed
a generic solar system model based on the \ac{BCR4BP}.
We showed how \ac{TST} approaches can be used to
elucidate the dynamics of satellites in the vicinity of
the sun--planet libration point L2,
and to determine the extent to which they are
influenced by the presence of a moon.

Two sets of parameters were investigated.
The first was chosen to show a more significant effect
while still resembling real, experimentally observed extrasolar systems.
Although there has not yet been a
confirmed moon observed around exoplanets~\cite{Teachey2018, Rodenbeck2018},
this work provides a basis for describing L2 for such systems in the future.
For comparison, the calculations were repeated
for parameters resembling our Solar System.
We found that a moon has a destabilizing effect on satellites near L2
when compared to a system in which planet and moon are merged into a single body.
Interestingly, the decay rate
when the planet and moon are collinear with the sun and L2
is lower than the average rate,
suggesting a reduced need for stabilization by a satellite at that instant.
To obtain these results,
the \ac{LMA} method has been generalized
to address velocity-dependent forces, such as the Coriolis force.
(See \ref{sec:lma} for details.)

We found that \ac{TST} can be used
to describe not only driven chemical reactions,
but also driven astrophysical systems.
The rate of instability calculated in this paper
describes how fast trajectories depart from the invariant manifold at L2
when they are not located exactly on said manifold.
In principle, this rate therefore relates to
the amount of fuel needed to actively stabilize a satellite on a specific orbit.
When using system parameters typical of our Solar System,
the methods presented here could be used to optimize the orbits of satellites
with respect to fuel consumption.


\section*{Declaration of Competing Interest}

The authors declare that they have
no known competing financial interests or personal relationships
that could have appeared to influence the work reported in this paper.


\section*{CRediT authorship contribution statement}

\textbf{Johannes Reiff:}
    Methodology,
    Software,
    Validation,
    Formal analysis,
    Investigation,
    Resources,
    Data Curation,
    Writing -- Original Draft,
    Writing -- Review \& Editing,
    Visualization,
    Supervision.
\textbf{Jonas Zatsch:}
    Software,
    Formal analysis,
    Investigation,
    Writing -- Original Draft.
\textbf{Jörg Main:}
    Conceptualization,
    Methodology,
    Formal analysis,
    Resources,
    Writing -- Original Draft,
    Writing -- Review \& Editing,
    Supervision,
    Project administration,
    Funding acquisition.
\textbf{Rigoberto Hernandez:}
    Conceptualization,
    Methodology,
    Writing -- Review \& Editing,
    Project administration,
    Funding acquisition.


\section*{Acknowledgments}

We thank A.~Junginger for initiating this project and M.~Feldmaier for
fruitful discussions.
The German portion of this collaborative work was partially supported
by Deutsche Forschungsgemeinschaft (DFG) through Grant No.~MA1639/14-1.
The US portion was partially supported by the National Science
Foundation (NSF) through Grant No.~CHE 1700749.
This collaboration has also benefited from support by the European
Union's Horizon 2020 Research and Innovation Program under the Marie
Skłodowska-Curie Grant Agreement No.~734557.


\appendix

\section{Jacobian of the equations of motion}
\label{sec:jacobian}

The second-order equations of motion
can be reformulated as a first-order system
\begin{equation}
    \dot{x} = v_x
    \qc
    \dot{y} = v_y
    \qc
    \dot{v}_x = \ddot{x}
    \qand
    \dot{v}_y = \ddot{y}
\end{equation}
with $\ddot{x}$ and $\ddot{y}$ given in Eq.~\eqref{eq:solar_system/eom}.
The Jacobian $\mat{J}$ of this first-order system reads
\begin{equation}
    \mat{J} = \mqty(
        0          & 0          &  1 & 0  \\
        0          & 0          &  0 & 1  \\
        J_{v_x, x} & J_{v_x, y} &  0 & 2  \\
        J_{v_y, x} & J_{v_y, y} & -2 & 0
    )
\end{equation}
with
\begin{subequations}
    \begin{align}
        J_{v_j, j}
            &= \pdv{\dot{v}_j}{j}
            = 1 + \sum_{k \in \set{\bS, \bE, \bM}} \qty[
                \frac{3 \mass{k}}{\dist{k}^5} (\vdist{k} \vdot \vunit{j})^2
                - \frac{\mass{k}}{\dist{k}^3}
            ]
        \qfor j \in \set{x, y}
        \\ \qand
        J_{v_x, y} = J_{v_y, x}
            &= \pdv{\dot{v}_x}{y}
            = \sum_{k \in \set{\bS, \bE, \bM}}
                \frac{3 \mass{K}}{\dist{k}^5}
                (\vdist{k} \vdot \vunit{x}) (\vdist{k} \vdot \vunit{y})
        \eqperiod
    \end{align}
\end{subequations}
It describes the linearized motion of satellites
relative to a reference trajectory.


\section{Local manifold analysis}
\label{sec:lma}

As in the ensemble method described in Sec.~\ref{sec:tst/instant_rate},
in the \ac{LMA}, we consider
the region of phase space close to a trajectory
$\gammaTS = (\xTS, \yTS, \pxTS, \pyTS)^\T$
on the \ac{NHIM},
as shown in Fig.~\ref{fig:ensemble}.
For simplicity, we choose a
moving coordinate frame in which the origin is at
$\xTS = \pxTS = 0$ for all times $t$.
The decay rate is determined by two components:

(i) The first contribution arises from the movement
of the ensemble akin to Sec.~\ref{sec:tst/instant_rate}
relative to \gammaTS.
The resulting flux through the associated \ac{DS} at $\xDS = 0$
(red diamonds in Fig.~\ref{fig:ensemble})
is then obtained via the slopes of the stable and unstable manifolds
defined by the variables $\xS = \xU$, \pxS, and \pxU.

(ii) The second contribution accounts for the fact that
in systems with more than one degree of freedom,
the ensemble can turn out of the $(x, p_x)$ plane associated with \gammaTS.
This can happen
if the system's orthogonal modes are coupled
to the reaction coordinate momentum $p_x$
and leads to an apparent movement of the \ac{DS} relative to \gammaTS\
(represented by the top arrow in Fig.~\ref{fig:ensemble}).
As a result, the instantaneous flux through the \ac{DS} is modified.
To quantify this effect,
we first propagate the top particle of the ensemble
initially located on the \ac{DS}
numerically by a small time step $\delta t$.
The related shift of the \ac{DS}, $\xDS(t + \var{t})$,
can then be determined by
projecting the propagated particle back onto the \ac{NHIM} using the \ac{BCM}.

Combining the two terms, the instantaneous decay rate can be written as
\begin{equation}
    \label{eq:lma/k}
    \kMani(t; \gammaTS)
    = \mat{J}_{x, p_x}(t) \frac{\pxU(t) - \pxS(t)}{\xU(t)}
        - \frac{\xDS(t + \var{t})}{\xU(t) \var{t}}
    \eqcomma
\end{equation}
where $\mat{J}(t)$ is the Jacobian of the system's equations of motion
evaluated for the trajectory \gammaTS\ at time $t$.

For a more detailed derivation of Eq.~\eqref{eq:lma/k} we consider a
trajectory $\gammaTS(t) = (\xTS, \yTS, \pxTS, \pyTS)^\T$
starting at some arbitrary time $t_0$ on the \ac{NHIM}.
All of the statements in this section
depend implicitly on \gammaTS\ and $t_0$,
which we neglect in our notation for simplicity.
Without loss of generality,
we choose coordinates such that $\xTS(t) = \pxTS(t) = 0$ for all times $t$.
Figure~\ref{fig:ensemble} sketches
an $(x, p_x)$-section of phase space in close proximity to $\gammaTS(t_0)$.
We can assume that the manifold fibers in this section are straight lines
since we only look at the dynamics very close to the \ac{NHIM}.
Therefore, the stable and unstable manifolds can be described
using only two vectors
$\gammaS = (\xS, \pxS)^\T$ and $\gammaU = (\xU, \pxU)^\T$.
These vectors will be squeezed and stretched as a function of time
if subjected to the equations of motion.
Without loss of generality,
we initially choose $0 < \xS(t_0) = \xU(t_0)$
such that we are in the linear regime.
In what follows, we reproduce the derivation of the \ac{LMA}
that was earlier included in Sec.~C
of the Supplemental Material of Ref.~\cite{hern20m}.

To obtain a decay rate for $\gammaTS(t_0)$, we now consider
a linear, equidistant ensemble parameterized by
\begin{equation}
    \label{eq:lma/ensemble}
    \gammaE(\alpha, t) = -\gammaS(t) + \alpha \gammaU(t)
\end{equation}
where $\alpha \in [0, 1]$.
The ensemble is constructed parallel to the
unstable manifold---see Fig.~\ref{fig:ensemble}.
Initially, the ensemble pierces the \ac{DS} at $\aDS(t_0) = 1$
(circles in Fig.~\ref{fig:ensemble}).
As time goes by, however, the ensemble will be stretched
and $\aDS(t)$ will therefore decay exponentially
(diamonds in Fig.~\ref{fig:ensemble}).
More precisely, \aDS\ is proportional to the number of reactants
and therefore leads to a decay rate
\begin{equation}
    \label{eq:lma/decay}
    \kMani(t_0)
    = -\eval{\dv{t} \ln(\aDS(t))}_{t_0}
    = -\aDSdot(t_0)
\end{equation}
at time $t_0$ in analogy to Eq.~\eqref{eq:tst/ensemble_rate}.
In this picture, the total decay rate consists of two contributions
\begin{equation}
    \label{eq:lma/parts}
    \kMani(t_0) = \kEns(t_0) + \kDS(t_0)
    \eqperiod
\end{equation}

For the first part, \kEns, we assume that
the ensemble stays in the $(x, p_x)$-section associated with $\gammaTS(t)$.
As a result, the point where the ensemble pierces the \ac{DS}
is fixed at $\xDS(t) = 0$ for all times $t$.
This is an effectively one-dimensional model.
We start by looking at the linearized dynamics near the \ac{NHIM}
\begin{equation}
    \dv{t} \vec{\gamma}(t) = \mat{J}(t) \vec{\gamma}(t)
    \eqcomma
\end{equation}
where $\mat{J}(t)$ is the Jacobian of the system's equations of motion
evaluated on the trajectory \gammaTS.
The fundamental matrix $\mat{\sigma}(t)$
obtained by integrating $\dot{\mat{\sigma}}(t) = \mat{J}(t) \mat{\sigma}(t)$
with $\mat{\sigma}(t_0) = \mat{1}_{2 d}$
can then be used to propagate the ensemble
from time $t_0$ to a later time $t$ via
\begin{equation}
    \gammaE(\alpha, t) = \mat{\sigma}(t) \gammaE(\alpha, t_0)
    \eqperiod
\end{equation}
We are interested in the point $\aDS$ where $\gammaE(\alpha, t)$ pierces the \ac{DS}
at $\xDS(t) = 0$,
\ie,
\begin{equation}
    \mat{\sigma}(t) \gammaE(\aDS(t), t_0) \vdot \vu{e}_x \overset{!}{=} 0
    \eqperiod
\end{equation}
Inserting the ensemble's parameterization defined in Eq.~\eqref{eq:lma/ensemble}
yields
\begin{equation}
    \aDS(t) = \frac
        {\mat{\sigma}_{x, x}(t) \xU(t_0) + \mat{\sigma}_{x, p_x}(t) \pxS(t_0)}
        {\mat{\sigma}_{x, x}(t) \xU(t_0) + \mat{\sigma}_{x, p_x}(t) \pxU(t_0)}
    \eqcomma
\end{equation}
where we have used $\xS(t_0) = \xU(t_0)$.
This intermediate result can be substituted into Eq.~\eqref{eq:lma/decay}.
Since we are only interested in the instantaneous rate at $t = t_0$,
we can simplify the expression
using $\mat{\sigma}(t_0) = \mat{1}_{2 d}$
as well as $\dot{\mat{\sigma}}_{x, p_x}(t_0) = \mat{J}_{x, p_x}(t_0)$
and arrive at
\begin{equation}
    \label{eq:lma/ens}
    \kEns(t_0)
    = \mat{J}_{x, p_x} \frac{\pxU(t_0) - \pxS(t_0)}{\xU(t_0)}
    \eqperiod
\end{equation}
A geometric interpretation of \kEns\ can be found in Ref.~\cite{hern19e}.

The second contribution, \kDS, in Eq.~\eqref{eq:lma/parts} stems from
the fact that in systems with more than one degree of freedom
the ensemble may leave the $(x, p_x)$-section associated with $\gammaTS(t)$.
An ensemble moving out-of-plane can mostly be treated as described above
by projecting it back onto the $(x, p_x)$-section.
Since the position of the \ac{DS} $\xDS(\vec{y}, \vec{p}_y)$
is dependent on the orthogonal modes, however,
this may lead to the ensemble intersecting with the \ac{DS} at $\xDS \ne 0$.

To quantify the effect on \kMani, consider a small time step $\var{t}$.
In the linear regime, the change $\var{\alpha}$ caused
by the ensemble drifting out-of-plane
can be written as
\begin{equation}
    \var{\aDS(t_0)}
    = \frac{\xDS(t_0 + \var{t}) - \xDS(t_0)}{\xU(t_0)}
    \eqcomma
\end{equation}
where $\xU(t_0)$ accounts for normalization.
Using Eq.~\eqref{eq:lma/decay} and the fact that $\xDS(t_0) = 0$, we obtain
\begin{equation}
    \label{eq:lma/ds}
    \kDS(t_0)
    = -\fdv{\aDS(t_0)}{t}
    = -\frac{\xDS(t_0 + \var{t})}{\xU(t_0) \var{t}}
    \eqperiod
\end{equation}
The quantities $\var{t}$ and \xU\ can be freely chosen within certain limits,
while $\xDS(t_0 + \var{t})$ can be determined numerically by
propagating the particle $\gammaE(1, t_0)$ initially located on the \ac{DS}
for $\var{t}$ units of time
and projecting it back onto the \ac{NHIM}
using the \ac{BCM}~\cite{hern18g}.
By combining the contributions~\eqref{eq:lma/ens} and~\eqref{eq:lma/ds}
according to Eq.~\eqref{eq:lma/parts},
we finally arrive at the instantaneous decay rate $\kMani(t; \gammaTS)$
for a trajectory \gammaTS\ on the \ac{NHIM} given in Eq.~\eqref{eq:lma/k}.


\bibliography{paper-q28}

\end{document}